\documentclass[preprint,aps,showpacs,floatfix,unsortedaddress]{revtex4}
\usepackage{graphics}

\begin{document}

\preprint{LA-UR-02-7047}
\title{Nonlinear Regge trajectories and glueballs}
\author{M.M. Brisudova}
\email{martina_brisudova@mentor.com}
\affiliation{Nuclear Theory Center, Indiana
University, 2401~Milo~B.~Sampson~Lane, Bloomington,~IN~47408}
\altaffiliation[On leave from: ]{Physics Institute,
Slovak Acad. Sci., D\'{u}bravsk\'{a} cesta 9,  842 28 
Bratislava, Slovakia}
\altaffiliation[\\ Present address: ]{Mentor Graphics Corporation, 
880 Ridder Park Drive, San Jose, CA 95131} 
\author{L. Burakovsky}
\email{burakov@lanl.gov}
\author{T. Goldman}
\email{tgoldman@lanl.gov}
\affiliation{Theoretical Division, MS B283, Los Alamos National
Laboratory, Los Alamos,~NM~87545}
\author{A. Szczepaniak}
\email{aszczepa@indiana.edu} 
\affiliation{Physics Department and Nuclear Theory
	Center, Indiana University, 2401~Milo~B.~Sampson~Lane, 
	Bloomington,~IN~47408}
\date{March 5, 2003}

\begin{abstract}
We apply a phenomenological approach based on nonlinear Regge
trajectories to glueball states. The parameters, i.e., intercept and
threshold, or trajectory termination point beyond which no bound states
should exist, are determined from pomeron (scattering) data.
Systematic errors inherent to the approach are discussed. We then
predict masses of glueballs on the tensor trajectory. For comparison,
the approach is applied to available quenched lattice data. We find a
discrepancy between the lattice based thresholds and the pomeron
threshold that we extract from data. 
\end{abstract}


\maketitle

\section{Introduction}

In previous work~\cite{us} we presented theoretical arguments and
strong phenomenological evidence that Regge trajectories of ordinary
quark-antiquark mesons are essentially nonlinear and can be well
approximated, for all practical purposes, by a specific function, the
so called square-root form. With a few additional assumptions intended
to reduce the number of independent parameters, and when possible,
tested for self-consistency, we obtained a remarkable agreement with
{\it both} bound (resonant) state ($t > 0$) {\it and} scattering 
($t < 0$) data. What makes this success even more impressive is the 
fact that the input parameters, with the exception of one, were 
determined by the masses of only a few lowest lying bound states.

The theoretical motivation for our previous study was the view that 
the properties of the gluon field, {\it e.g.} the flux tube, change 
with the increasing size of the hadron. At long enough distances, a 
linear potential is simply a wrong approximation to the interaction 
of meson constituents. Therefore, Regge trajectories cannot be 
asymptotically linear~\cite{ftnt} in $t$, even if they appear to be 
so over a limited range of $t$. In searching for a more practical 
approximation, we studied nonlinear Regge trajectories corresponding 
to dual amplitudes with Mandelstam analyticity (DAMA)~\cite{DAMA} in 
a simplified situation, that is, toy models which, nevertheless, 
maintain a resemblance to QCD~\cite{us,LEO,BBG}.  After thorough 
investigation, we argued that (i) DAMA amplitudes can be expected to 
fit spectra; (ii) nonlinearity arises due to the color screening of 
the flux tube, and thus, (iii) one can expect the same qualitative 
behavior of the trajectories regardless of quantum numbers of the 
quark-antiquark meson.

In this paper we attempt to go beyond ordinary quark-antiquark mesons
to trajectories for glueballs. However, it is not {\it a priori} clear
that non-$\bar{q}q$ mesons such as glueballs or hybrids can be 
satisfactorily described by the same form of Regge trajectories as 
are ordinary mesons. Even though the nonlinearity of Regge trajectories is 
due to color screening, which is not exclusive to $q\bar{q}$ mesons, the
behavior of glue in exotic systems can be different.  In particular, in
the case of glueballs there have been contradictory opinions about the
basic nature of these states, ranging from solitonic through loops of
glue to glue strings. The latest lattice QCD results indicate that
the loops-of-glue picture does not agree with the lattice spectra to
the extent that the bag model does, thus supporting a constituent-like
picture for gluons (in addition to quarks)~\cite{colin}. (However, 
it should be kept in mind that the flux tube model is, by definition,
a very simple model with a big symmetry group and so cannot be used to 
make detailed predictions for the spectrum; the flux tube-like 
distribution of the fields should be measured instead.) The same
authors view hybrids as states where a quark and an antiquark are
subject to a potential corresponding, roughly speaking, to an excited
flux tube. They generate a plethora of these potentials in quenched
QCD. Obviously, in unquenched QCD the picture can be drastically
different.  It is, unfortunately, challenging and difficult at present
to capture the physics of string breaking accurately with lattice
methods~\cite{colin}. Here we focus on pure gluonic states and only
comment on the prospect of extending our analysis to hybrid mesons in
the conclusion.

Encouraged by the lattice QCD analysis of Ref.\cite{colin}, we apply
the same phenomenological approach that worked so remarkably well for
ordinary mesons to the pomeron trajectory.  In a departure from our
previous calculations, we analyze scattering data to fit parameters of
the square-root Regge trajectory. Naturally, in this case the extracted
value of the threshold, or termination point of the real part of the
trajectory, is subject to  a large uncertainty. To reduce the
uncertainty, we add to our data set the mass of the lowest lying tensor
glueball determined in Ref.\cite{colin}.  

%%%%%%%%%%%%%%%%%%%%%%%%%%%%%%%%%%%%%%%%%%%%%%%%%%%%%%%%%%%%%%%%%%%%% 
(There are a number of other lattice QCD determinations of this mass, 
see e.g.~Refs.\cite{others}, but the rms variation of the values 
reported is only about 1\% which is considerably smaller than the 
uncertainties in the individual determinations. Hence, choosing 
one specific case has a negligible effect on our results compared 
to the overall (statistical and systematic) uncertainties below 
and avoids a perhaps unrealistic reduction of the total uncertainty.) 
%%%%%%%%%%%%%%%%%%%%%%%%%%%%%%%%%%%%%%%%%%%%%%%%%%%%%%%%%%%%%%%%%%%%%%

						With this addition, the
uncertainties in the trajectory parameters are significantly reduced
while the values are little affected. We take the value of the
threshold as an indication of a maximum mass beyond which no glueball
states exist, bearing in mind that we are unable to prove at present
that the square-root form will be as comparably efficient in describing
glueballs as it was for ordinary mesons.

We should also note that the threshold we find here is larger than 
what is inferred from trajectories fitted to quenched lattice QCD 
glueball mass states. Whether this is due to the difference between 
quenched and unquenched QCD, or the functional forms used for the 
trajectories remains unknown. 

The choice of a specific trajectory within the allowed range of DAMA
trajectories introduces an unknown systematic error. In case of
ordinary mesons, our results justified the assumptions {\it a
posteriori}.  Due to insufficient data, this is not possible when
dealing with pure glue trajectories. To estimate the systematic errors,
we repeat the fit with the other limiting form of DAMA trajectories,
the so called logarithmic form. We take the difference between the
thresholds obtained is indicative of the systematic errors.

While the pomeron trajectory is the only glueball trajectory for which
unambiguous data exist, it may be, unfortunately, affected by the gluon
condensate. In the absence of experimental glue bound state data, we
use glueball masses from lattice QCD to determine the thresholds of
other, less peculiar trajectories, bearing in mind that the masses are
subject to the quenched approximation.

This paper is organized as follows: Section II starts with a brief
introduction to the DAMA trajectories, followed by our fits to pomeron
data. We compare the fits for the two limiting forms of trajectories,
and discuss the physical meaning of the results. Section III is devoted
to glueball spectroscopy. We conclude with a comment on possible future
applications.

\section{DAMA trajectories and fits to data} 

The class of dual models called dual amplitudes with Mandelstam
analyticity (DAMA)~\cite{DAMA} is a generalization of Veneziano
amplitudes~\cite{Ven} to the most general form consistent with
Mandelstam analyticity.  (DAMA has the Veneziano limit $\alpha (t)\sim
t,$ but the transition to this limit occurs
discontinuously~\cite{LASZLO}.)

A meson trajectory $\alpha _{j\bar{i}}(t),$ can be parametrized on the
entire physical sheet in the following form:
\begin{eqnarray}
\alpha _{j\bar{i}}(t)=\alpha _{j\bar{i}}(0)+\gamma_{\nu} \Big[
T_{j\bar{i}}^\nu -(T_{j\bar{i}}-t)^\nu \Big] ,\;\;\;  0 \leq \nu \leq
{1\over{2}}. \label{damanu}
\end{eqnarray}
(up to a power of a logarithm), assuming that: $\alpha_{j\bar{i}} (t)$
is an analytic function having a physical cut from some value $t_0$ to
$\infty$; it is polynomially bounded on the entire physical sheet, and
there exists a finite limit of the trajectory phase as $|t|\rightarrow
\infty $ \cite{Tru}.  The subscripts $i,j$ indicate dependence of the
parameters on the  flavor content of the meson within a meson
multiplet. In this paper  we drop the subscripts for simplicity.

The parameter $\gamma_{\nu} $ is the universal asymptotic slope for
nonlinear trajectories \cite{CJ}, $$\alpha (t)\sim -\gamma_{\nu}
(-t)^\nu,\;|t|\rightarrow  \infty ;$$ both $\gamma_{\nu}$ and the
exponent $\nu$ are independent of quantum numbers. In order for the
slope to be positive at small $t$, $\gamma_{\nu}>0$.  The intercept
$\alpha_{j\bar{i}}(0)$  varies for different trajectories, and in
accord with the Froissart bound should satisfy $\alpha_{j\bar{i}}(0)
\leq 1$. In reality, however, there is an exception --- the intercept
of the pomeron trajectory is observed to be slightly larger than 1. The
parameter $T_{j\bar{i}}$ is often called the trajectory threshold.

Note that for $\vert t \vert \ll T,$ Eq. (\ref{damanu}) reduces to a
(quasi)linear form:
\begin{eqnarray}
\alpha _{j\bar{i}}(t) = \alpha _{j\bar{i}}(0) + \nu \gamma 
T_{j\bar{i}}^{\nu -1}t = \alpha _{j\bar{i}}(0) +
\alpha'_{j\bar{i}}(0)\;\!t.                   \label{nonl}
\end{eqnarray}

The value of $\nu$ is restricted to lie between $0$ and $1/2$, in
accordance with Ref.~\cite{LASZLO}. The value $\nu=0$ should be
understood as a limit $\nu \rightarrow 0$, $\gamma_{\nu} \nu $ fixed.
In this limit, the difference of fractional powers reduces to a
logarithm, viz.,
\begin{eqnarray}
\alpha (t)=\alpha (0)- \gamma_{\log}  \log \left(
 1-\frac{t}{T_{\log}}\right)
,\;\;\;
\gamma_{\log} \equiv \lim_{\nu \rightarrow 0} \gamma_{\nu} \nu
\label{log}
\end{eqnarray}
Unlike a trajectory with any value of $\nu \neq 0$, the real part of
the ``logarithmic'' trajectory does not freeze-out when $t$ reaches
$T$. The real part continues to grow; the only change  for $t>T$ is
that the trajectory acquires a constant imaginary part.

The upper bound on $\nu$ gives the so-called ``square-root''
trajectory, viz.
\begin{eqnarray}
\alpha (t)=\alpha (0)+\gamma_{1/2} \Big[ \sqrt{T} -\sqrt{T-t} \Big] .
\label{sqrt}
\end{eqnarray}
When $t$ reaches $T$, the real part of the ``square-root'' trajectory
stops growing, and there are no states with a higher angular momentum
than $ \ell_{max} =\Big[ \alpha (T)\Big]$. For this reason, the
parameter $T$ is also called the trajectory termination point. This is
true for any value of $\nu \neq 0$.

\subsection{Fit with the square root form}

In Ref.\cite{us} we used only the square-root form of Eq.(\ref{sqrt})
for spectroscopy purposes. We determined the value of $\gamma_{1/2}$
from the $\rho$  trajectory, and then it was taken as universal for all
other meson  trajectories.  Our calculation is in excellent agreement
with various data, not only spectroscopic but scattering as well, and
is self-consistent, justifying the assumptions {\it a posteriori}.
This leads us to believe that the extracted  value of $\gamma_{1/2}$ is
reliable, and that the square-root form is close to the  true
functional form of meson trajectories.
\begin{figure}
\includegraphics{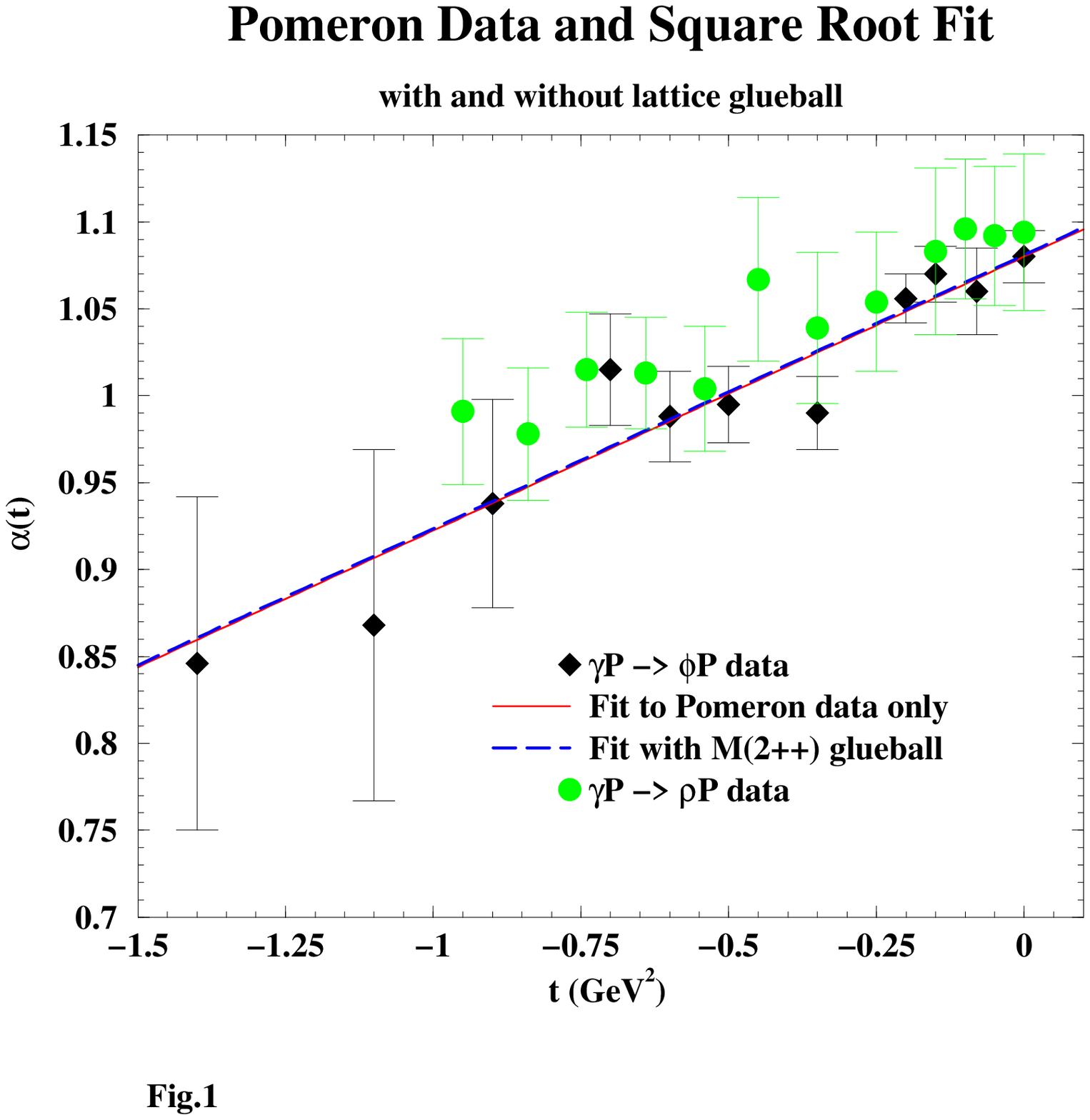}
\vskip1cm
\caption[]{Pomeron data together with our best fits of the square-root
form: The dashed line fit includes the averaged mass $M(2^{++})$ as a
data point; for the solid line fit only the scattering data was used.}
\end{figure}
Since $\gamma_{\nu}$ has to be a constant {\it independent} of the
flavor content or quantum numbers of the meson, we apply the same value,
\begin{equation}
\gamma_{1/2} =3.65\pm 0.05\;{\rm GeV}^{-1},
\end{equation} 
\noindent to the data for the pomeron.

With the value of the universal asymptotic slope fixed, we can now fit
the pomeron scattering data to find the remaining parameter of the
pomeron trajectory, $T_{1/2}$. Of the two data sets~\cite{data} shown
in Fig.~1, we use only $\gamma p \rightarrow \phi p $ since at ZEUS 
energies, it is devoid of significant contributions from exchanges 
other than the pomeron.  For comparison, we show both sets of data. 
Note that the data points corresponding to $\gamma p \rightarrow \rho^0 
p $ are consistently above the $\gamma p \rightarrow \phi p $ as 
expected due to contributions from additional exchanges.

In Fig.~1, the solid line shows our best fit of the square-root form,
Eq.(\ref{sqrt}):
\begin{eqnarray}
\alpha_{1/2 }(0)& = &1.08 \pm 0.01,  \\
\sqrt{T_{1/2}} & = & 11.56 \pm 2.08 {\rm GeV} \label{TalphaBez} \\
& \chi^2/{\rm d.o.f.} = 6.07/9 . \nonumber
\end{eqnarray}

The relatively large, 18\% error on the fitted threshold is caused in
part by the data uncertainties, and in part by the fact that all data
points are concentrated in a small region of $t$ near zero; 
consequently,  $\vert t \vert << T$.  The fit is essentially dominated
by the first term in the Taylor expansion of Eq.(\ref{sqrt}):
\begin{eqnarray}
\alpha_{1/2} (t)& = & \alpha(0) + {\gamma_{1/2}\over{2\sqrt{T_{1/2}}}}\,
t \label{alfalinsqrt} \\
& = & 1.08 + 0.1578 \ [{\rm GeV}^{-2}] \ t
\end{eqnarray}

In order to determine the threshold with a better accuracy, additional
data is needed. We use the mass of the lowest lying tensor glueball
from Ref.\cite{colin}, 
$M(2^{++}) = 2.40 \pm 0.13$~GeV. With this additional data  point, the
error on the extracted threshold is reduced by factor of two, while the
$\chi^2$ of the fit remains comparably small and the fitted parameter
values are essentially unchanged. (See Figure 1). In this way we
obtain:
\begin{eqnarray}
\alpha_{1/2 }(0)& = &1.081 \pm 0.007, \\
\sqrt{T_{1/2}} & = & 11.57 \pm 1.1 {\rm GeV} \label{Talpha}
\\ &  \chi^2/{\rm d.o.f.} = 6.07/10 . \nonumber
\end{eqnarray}

We note that the $\chi^2/{\rm d.o.f.}$ in both cases is smaller than
one would expect on general grounds. This is similar to the case
observed in Ref.\cite{AP} for other trajectories, presumably for
similar reasons.

\subsection{Fit with the logarithmic form}

Even though the logarithmic trajectory, Eq.(\ref{log}), itself is not
realistic, since  its  real part grows without bound (in contrast to
any other trajectory of our nonlinear form, Eq.(\ref{nonl}), with $\nu
\neq 0$), it is useful to study because the true trajectory can lie
anywhere between the two limiting forms. In addition to this reason,
comparison of the fits with the two limiting forms can, to some extent,
illuminate the issue of systematics.  There is undoubtedly a systematic
error associated with the choice of DAMA trajectories as the class of
trajectories within which the true trajectory lies, and this we cannot
estimate. There is an additional systematic error arising from the
specific choice of $\nu=1/2$ within the model and  from the way we
determine the parameters of the trajectories. It is this second
uncertainty that we address in this section.

To fit the data with a logarithmic form, we first need to determine the
value of $\gamma_{\log}$.  In complete analogy with our calculation
utilizing the square root form, the value of $\gamma_{\log}$ is
determined from the $\rho$ trajectory.  We find
\begin{eqnarray}
\gamma_{\log} = 8.00\pm 0.34\; .
\end{eqnarray}
\noindent
Note that $\gamma_{\log}$ is dimensionless.

Our best fit of the logarithmic form, Eq.(\ref{log}), to the scattering
pomeron data,
\begin{eqnarray}
\alpha_{\log}(0) & = & 1.08 \pm 0.01,  \\
\sqrt{T_{\log}} & = & 7.09 \pm 0.64 {\rm GeV} \\
 &   \chi^2/{\rm d.o.f.} = 6.07/9 \nonumber
\end{eqnarray}
is shown in Fig.~2. Also in Fig.~2, we show the fit to the scattering
data plus mass of the tensor glueball, yielding
\begin{eqnarray}
\alpha_{\log}(0) & = & 1.079 \pm 0.007,  \\
\sqrt{T_{\log}} & = & 7.23 \pm 0.34 {\rm GeV} \\
 &   \chi^2/{\rm d.o.f.} = 6.12/10 .  \nonumber
\end{eqnarray}
The addition of the bound state data point leads to a $2\%$ increase in
$\sqrt{T_{\log}}$ and almost a factor of two reduction in its error.
Recall that, for the square root form, the threshold remains almost the
same (less than $0.1\%$ change).

Again, not surprisingly, the fit is dominated by the linear term
\begin{eqnarray}
\alpha_{\log}(t)& = & \alpha(0) + {\gamma_{log}\over{T_{log}}}\, t
\label{alfalinlog} \\  &= & 1.08 + 0.1590 \ [{\rm GeV}^{-2}]\ t .
\end{eqnarray}

\subsection{Estimate of systematic error}

The linear term dominance in all of our fits implies that the
thresholds of the square root and logarithmic trajectories, extracted
from pomeron data, are simply related.  The relative size of the
threshold for the two limiting forms follows directly from comparison
of the two linearized trajectories, Eq.(\ref{alfalinsqrt}) and
Eq.(\ref{alfalinlog}):
\begin{eqnarray} 
T_{\log} = 2 {\gamma_{\log}\over{\gamma_{1/2}}}\sqrt{T_{1/2}} . 
\end{eqnarray}

The ratio of ${\gamma_{\log}\over{\gamma_{1/2}}}$ is, in our
calculation, fixed by the $\rho$ trajectory data used as input.  To
find the three parameters of the $\rho$ trajectory, (its intercept,
threshold and the universal parameter $\gamma_{\nu}$ for any chosen
$\nu$), we restrict  the trajectory to pass through the three
experimentally well established points, specifically the intercept, and
the mass and spin of the $\rho$ and the $\rho_3$.

The straight line that crosses $\rho$ and $\rho_3$ leads to an intercept
smaller that the observed value. This means that the linear form is
insufficient~\cite{us}. By an {\it a posteriori} comparison of the DAMA
trajectory to its truncated Taylor series, one can see that the fit is
basically dominated by terms up to ${\cal O}(t^2)$, viz.
\begin{eqnarray}
  \tilde{\alpha}_{1/2}(t) &\simeq & \tilde{\alpha}(0)
+{\gamma_{1/2}\over{2 \tilde{T}_{1/2}^{1/2}}}\, t +
{\gamma_{1/2}\over{8 \tilde{T}_{1/2}^{3/2}}}\, t^2 ,  \label{sqo2}\\
  \tilde{\alpha}_{\log}(t)& \simeq & \tilde{\alpha}(0)
+ \  {\gamma_{\log}\over{\tilde{T}_{\log}}\ }\, t
 + \   {\gamma_{\log}\over{2\, \tilde{T}_{\log}^2}}\, t^2 .
\label{logo2} \end{eqnarray}
(We use the tilde to distinguish the $\rho$ trajectory from the pomeron
trajectory discussed so far.)  For example, the square root trajectory
evaluated at the mass of the $\rho_3$ differs from its Taylor series,
Eq.(\ref{sqo2}), by less than $0.5$\%.

Setting the Taylor series coefficients in Eqs.(\ref{sqo2}) and
(\ref{logo2}) equal, we obtain
\begin{eqnarray}
\tilde{T}_{\log} & \simeq & 2 \tilde{T}_{1/2} \label{rhoT} \\
\gamma_{\log} & \simeq & \gamma_{1/2} \sqrt{\tilde{T}_{1/2}}
\label{rhog}
\end{eqnarray}
This leads the following relation for the pomeron thresholds:
\begin{eqnarray}
T_{\log} \simeq 2 \, \sqrt{\tilde{T}_{1/2}} \, \sqrt{T_{1/2}}.
\label{tl} 
\end{eqnarray}
The numerical values we find are in a good agreement with these
relations.

Is there anything deep about these relations?  Note that
Eqs.(\ref{rhoT}) and (\ref{rhog}) are a direct consequence of the fit
being dominated by up to quadratic terms in $t$ and our requirement
that the trajectory pass {\it exactly} through the three input points.
Thus, Eq.(\ref{rhog}) cannot hold for a universal $\gamma$ unless all
thresholds are identical. Alternatively, the three point restriction is
too strong and/or only one specific value of $\nu$ can be correct.

\subsection{Understanding the systematic error in terms of toy models}

To understand this issue further, we turn to our toy models~\cite{us}.
Within the framework of our generalized string model, it is possible to
reconstruct the potential from a Regge trajectory~\cite{us,LEO}.
Earlier we found that potentials corresponding to the square-root and
to the logarithmic trajectories, respectively, are very close, when
normalised to the same asymptotic value. Furthermore, they are also
very close to a potential found from a fit to lattice data~\cite{Born}
that we used in another toy model, which consists, basically, of  a
leading order Born-Oppenheimer (LOBO) approximation for a system of a
very heavy quark and antiquark. In that toy model, we found that the
spectrum could be equally well fitted by both limiting forms of the
Regge trajectories, with nearly the same thresholds~\cite{us}.

The difference between the toy model study and the situation at hand is
the following: In the toy model we {\it fit} the data with many points,
in effect optimizing parameters of the underlying potentials so that
they produce  {\it similar} results  for a large number of bound
states. In the fit to real data, we  have instead a very few  lowest
lying bound states, and {\it solve} for the parameters of trajectories.
Implicitly, we demand that the underlying potentials produce the {\it
same} results at those input points. Since the points are the lowest
lying bound states, this corresponds to ``aligning'' the potentials in
the region relevant for the lowest lying bound states, which can, and
should be expected to, lead to different asymptotic values.  Since the
asymptotic value of the potential is directly related to the threshold
of the Regge trajectory (at least in the toy model), this translates
into the  thresholds for square root and logarithmic trajectories being
different.

The difference between the extracted thresholds is thus an indicator of
systematic errors. Combining errors in quadrature, we conclude that the 
threshold for the pomeron trajectory is ($9.4 \pm 1.6$) GeV.
\begin{figure}
\includegraphics{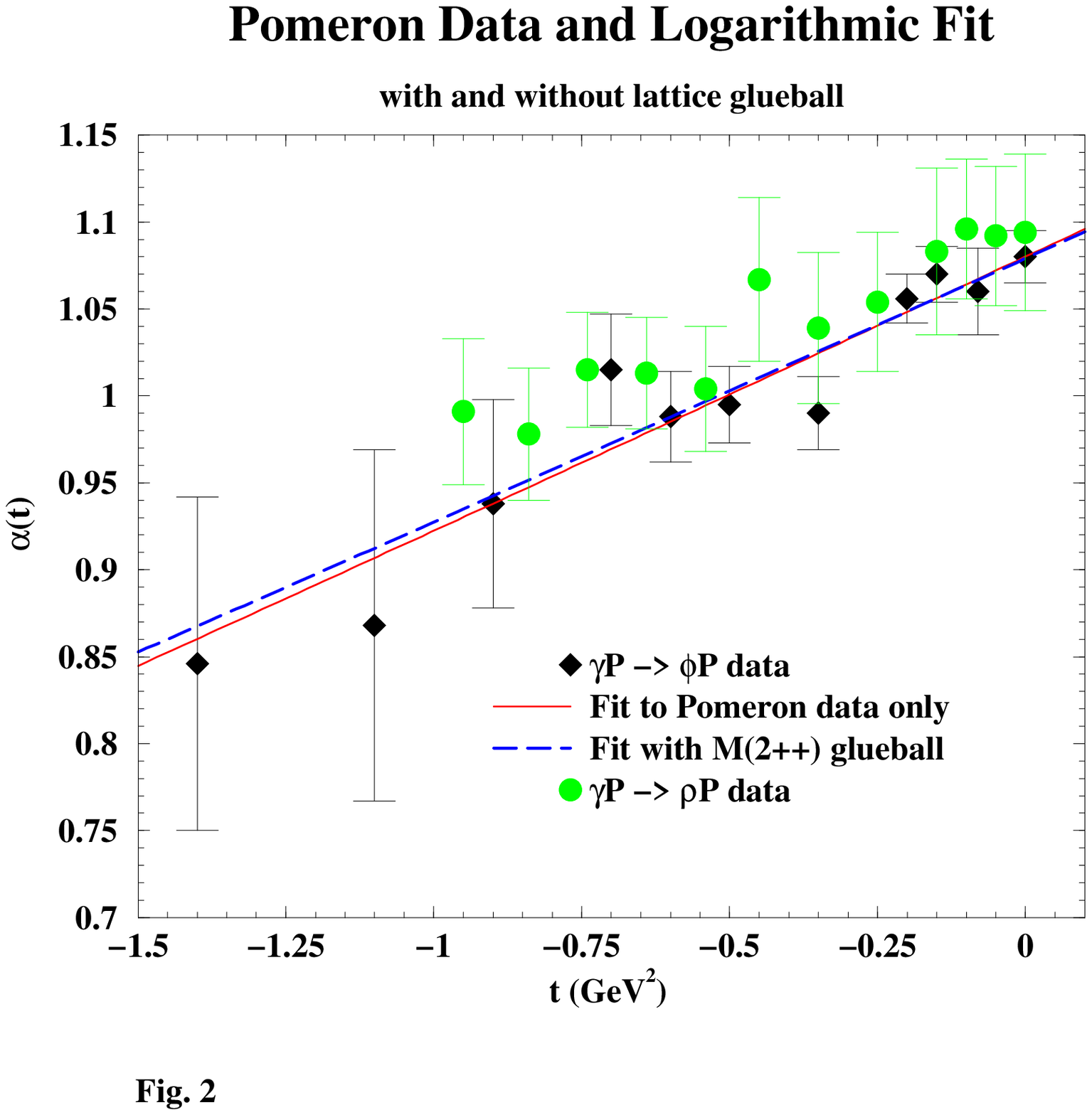}
\vskip1cm
\caption[]{Pomeron data together with our best fits of the logarithmic
form: The dashed line fit includes the averaged mass $M(2^{++})$ as a 
data point; for the solid line fit only the scattering data was used.}
\end{figure}

\section{Conclusions on Glueball Spectroscopy}

As is obvious from Eqs.(\ref{TalphaBez}) and (\ref{Talpha}), the square
root form of the trajectory with the parameters fitted to scattering
data alone gives the same mass predictions as the fit to both the
scattering data and the tensor glueball mass, but with larger errors.
The fit to both the scattering data and the mass of the $2^{++}$
glueball from lattice may give quite precise predictions for higher
excited states, providing the lattice value is close to the true mass
of the glueball. Note that our method works very well even for higher
excited states.  Using the fit, Eq.(\ref{Talpha}), we  obtain the
following predictions for excited glueball masses: $M(4^{++})= 4.21 \pm
0.21$~GeV, $M(6^{++}) = 5.41 \pm 0.28$~GeV, and we obtain $M(2^{++})=
2.38 \pm 0.12$~GeV, with  the same central value obtained from purely 
scattering pomeron data.

Based on our calculation we conclude that the threshold for tensor
glueballs can be expected in the region no lower than 7~to~8~GeV and no
higher than 11~to~12~GeV. From the fitted values of the thresholds for
various meson multiplets, we know that the thresholds for the same
flavors, but different multiplets, vary by less than 20\%. We expect
the same to be true for glueballs.

Note that glueball threshold is much larger than what we found for
ordinary ($q\bar{q}$) light mesons. This is related to the smaller
(local in $t$, not asymptotic, of course) slope for the pomeron 
trajectory but also devolves from our DAMA approach. From this higher 
value, we infer that a significantly larger number of higher states 
are available for glueballs than for mesons.

It is also interesting to contemplate what the maximal value of
$J=J_{\rm max}$ may be for the states allowed. We recall from our
previous work~\cite{us} that the square root trajectory tends to
overestimate the growth near the threshold; that is, at any given mass
it tends to predict slightly high angular momentum for states
approaching the threshold. This was expected from model studies and
further confirmed by fits to light meson spectra where sufficient data
were available. For example, for the $a_2$ trajectory the square root
form allowed for a $J=8$ state, whereas we concluded that $J=6$ should
actually be the last state on the trajectory.

The glueball trajectory allows for very large values of $J_{\rm max}$,
possibly as high as 36. The specific value of $J_{\rm max}$ is
obviously subject to a large uncertainty, but our conclusion is firm
that the value will be significantly larger than that for ordinary
mesons, possibly even larger than that for heavy quarkonia.  This
raises the paradoxical possibility that high $J$ glueballs may be
sufficiently narrow to identify experimentally. (The only other states 
expected to be available to mix with them would be heavy quarkonia. 
Such mixing is suppressed to perturbative values by the heavy quark 
mass.)

Unfortunately, the pomeron trajectory is the only glueball 
trajectory for which experimental data is available, and we 
expect that there are certain exceptional aspects associated 
with it: As for the light quark-antiquark system, a perturbative 
analysis shows a strong attraction between two gluons with a 
threshold at zero. Relativistically, this suggests that the 
two particle bound state will develop a negative mass-squared, 
requiring~\cite{NJL} the formation of a gluonic vacuum condensate 
(which is known to occur) and a mass gap for the lowest scalar 
state above the shifted vacuum. The pomeron trajectory does pass 
through $J=0$ at a negative mass-squared and there is, of course, 
no physical state there.  If the lowest mass scalar glueball is 
indeed the scalar state as shifted by the formation of the gluon 
condensate, it need not appear at the mass expected from standard 
consideration of the daughter trajectories of the pomeron.

Such a distortion, however, does not appear to occur~\cite{AP} for 
the $f_{0}(980)$ trajectory in the case of light quarks, where the 
issues of `four quark' states, as well as mixing (including with 
the scalar glueball), also arise. However, one does not know if this 
apparent regularity will hold for the pomeron trajectory states as 
well. Therefore, it is also interesting to investigate additional 
glueball states.

In the absence of other data, we turn to lattice QCD. So far, only the
spectrum of glueballs in quenched QCD has been calculated. Of the
states listed in Ref.\cite{colin}, the $0^{-+}$ at $2.59\pm0.17$ GeV and
$2^{-+}$ at $3.10\pm0.18$ GeV should form a common trajectory with an
intercept $-3.58\pm1.48$ and threshold $\sqrt{T}=3.90 \pm 0.91$ GeV.
Another trajectory is formed by a $1^{+-}$ at $2.94\pm 0.17$ GeV and a
$3^{+-}$ at $3.55 \pm 0.21$ GeV. Its intercept is $-2.60 \pm 1.45$ and
threshold $\sqrt{T}=4.87 \pm 1.25$ GeV. Note the large errors of the
extracted values.  The remaining data from Ref.\cite{colin} are
ambiguous for our purposes because they can contain admixtures of
higher angular momentum states. 

The values of thresholds extracted from lattice data are significantly
lower than the threshold of the pomeron trajectory, although, within
the large errors, they are consistent with the lower limit of the
pomeron threshold. Moreover, since the thresholds for different
trajectories of the same flavors can be expected to differ as much as
20\%, the apparent disagreement is not alarming. Unquenched lattice 
data, therefore, could be very useful for further advancing our 
understanding the nature of the pomeron trajectory, in the absence 
of further experimental evidence.

\section{Future prospects}

As a prospect for future work, we comment on the possibility of
applying our approach to hybrid mesons, which are of current interest
to both experiment and theory. In Ref.\cite{colin}, the authors used
the LOBO approximation to calculate spectra of bottomonium as well as
bottomonium-like hybrids. The LOBO approximation was demonstrated to be
an efficient and reliable method.  However, the potentials used as
inputs are generated in quenched lattice QCD, and thus, not all of the
states predicted may survive in an unquenched theory. Color screening
due to light quarks can be expected to become more and more important
with increasing size of the hadron. 

Exactly how many of the hybrid states can be reliably extracted from 
a quenched calculation is unclear. For example, the wave function of 
the lowest lying hybrid is found to be larger than that of the lowest 
lying quarkonium, but still smaller than the scale where flux tube 
breakage is expected~\cite{colin}.  There is evidence that survival 
of the lowest lying hybrids as well-defined resonances remains 
conceivable~\cite{hybrid}. 

Unfortunately, implementing the physics of flux tube breaking in 
lattice QCD is very difficult at present. This is where our
phenomenological approach can be of assistance. The key assumption is
that because the curvature of the Regge trajectory arises due to the
screening and breakage of flux tube, we can use the same (that is,
square root) form of the trajectories as for ordinary mesons.

It is also unfortunate that, at present only spin averaged lattice 
data are available for the bottomonium-like hybrids~\cite{colin}. If 
the physical states are nearly degenerate, i.e., spin splittings are 
small, then the spin averaged data can be used to extract the parameters 
of the Regge trajectories, and, in particular, the threshold value can 
be reliable. For example, in pure $q\bar{q}$ mesons, the bottom mass 
provides a sufficient suppresion factor for the subleading splittings, 
whereas the charm mass is insufficient.  In case of bottomonium-like 
hybrids, most possible spin dependent operators are suppressed by the 
heavy quark mass. However, there is a contribution due to the angular 
momentum of the glue that can be expected to be of the same magnitude 
as the LOBO splittings. This invalidates any conclusions one could 
draw using our approach based on the leading order splittings at 
present. Given more precise lattice data, it would be interesting to 
compare trajectory parameters of hybrids to those of ordinary mesons 
and or glueballs. 

It should also be worthwhile to examine daughter trajectories using our
approach. Conversely to the above, a plethora of relevant experimental 
evidence is available.  A key question here is whether the thresholds for 
the daughter trajectories are consistent with those found for the parents.

This research is supported by the Department of Energy under contracts
W-7405-ENG-36 and DE-FG02-87ER40365.


\begin{thebibliography}{9}
\bibitem{ftnt} Linear trajectories are, in fact, disfavored by various 
	experimental data. For more details, we refer the reader to 
	discussions in Ref.\cite{us}.
\end{thebibliography}
\end{document}